          [2004/06/22 1.2 (KOF); 2001/04/25 1.1 (PWD)]

\documentclass[a4paper,twocolumn]{Gaia2004} 
\usepackage{times}      
\usepackage{epsfig}     
\usepackage{natbib}     
\def\kms{km~s$^{-1}$}
\title{Dynamical streams in the solar neighbourhood}

\author{B. Famaey}
\author{A. Jorissen}
\affil{Institut d'Astronomie et d'Astrophysique, Universit\'e Libre de Bruxelles, Bvd du Triomphe, 1050 Bruxelles, Belgium}
\author{X. Luri}
\affil{Departament d'Astronomia i Meteorologia, Universitat de Barcelona, Avda. Diagonal 647, 08028 Barcelona, Spain}
\author{M. Mayor}
\author{S. Udry}
\affil{Observatoire de Gen\`eve, Chemin des Maillettes 51, 1290 Sauverny, Switzerland}
\author{H. Dejonghe}
\affil{Sterrenkundig Observatorium, Universiteit Gent, Krijgslaan 281, 9000 Gent, Belgium}
\author{C. Turon}
\affil{Observatoire de Paris-Meudon, 92195 Meudon cedex, France}

\bibpunct{(}{)}{;}{a}{}{,}  

\begin{document}


\maketitle

\begin{abstract}

The true nature of the Hyades and Sirius superclusters is still an open question. In this contribution, we confront Eggen's hypothesis that they are cluster remnants with the results of a kinematic analysis of more than 6000 K and M giants in the solar neighbourhood. This analysis includes new radial velocity data from a large survey performed with the CORAVEL spectrometer, complemented by Hipparcos parallaxes and Tycho-2 proper motions\ (Famaey et al. 2004). A maximum-likelihood method, based on a bayesian approach, has been applied to the data, in order to make full use of all the available data (including less precise parallaxes) and to derive the properties of the different kinematic subgroups. Two such subgroups can be identified with the Hyades and Sirius superclusters.  Stars belonging to them span a very wide range of age, which is difficult to account for in Eggen's scenario. These groups are thus most probably {\it dynamical streams} related to the dynamical perturbation by spiral waves rather than to cluster remnants. An explanation for the presence of young clusters  in the same area of velocity-space is that they have been put there by the same wave. Indeed, the scale of a cluster is small in comparison to the scale of the perturbation of the potential linked with a spiral wave. The response of a cluster to a spiral wave is thus similar to the response of a single star. Thus, in this scenario, the Hyades and Ursa Major clusters just {\it happen to be} in the Hyades and Sirius streams, which are purely dynamical features that have nothing to do with the remnants of more massive primordial clusters. 

This mechanism could be the key to understanding the presence of an old metal-rich population, and of many exoplanetary systems in our neighbourhood. Moreover, a strong spiral pattern seems to be needed in order to yield such prominent streams. Since spiral structure is usually baryonic, this would leave very little room for dark matter. 
This may be an indication that the era of the dark-matter paradigm explaining the dynamics of the Galaxy may come to an end, and is being superseded by modified gravity.

\end{abstract}

\section{Introduction}

It is known for a long time that spatially unbound groups of stars in the 
solar neighbourhood share the same kinematics as well known open clusters
(namely the Hyades and Ursa Major clusters; Eggen 1958, 1960). Assuming
that they are vestiges of more massive primordial clusters which partly
evaporated with time, those kinematically cold groups are generally
called {\it superclusters} (namely the Hyades and Sirius superclusters).
They are also often called {\it streams} or {\it moving groups}. 

During the last fifteen years, Eggen's hypothesis that these groups are 
in fact cluster remnants has been largely debated, because they may also
be generated by a number of global dynamical mechanisms. Kalnajs\ (1991)
suggested that a rotating bar at the centre of the Milky Way could cause
the velocity distribution in the vicinity of the outer Lindblad resonance
to become bimodal, due to the coexistence of orbits elongated along and
perpendicular to the bar's major axis. He presented this effect as a
possible explanation for the presence of superclusters in the solar
neighbourhood. Today, we know that this mechanism cannot account for the
Hyades and Sirius superclusters, but does account  for the Hercules
stream (Dehnen 2000), a group of stars lagging behind the galactic
rotation and moving outward in the disk. Nevertheless, the Hyades and
Sirius superclusters could still be linked with the spirality of the
Galaxy, because any perturbation of the axisymmetric potential is likely
to buffet the stars. The correct scenario to interpret the supercluster
phenomenon is thus still an open question. In this contribution, we try
to answer it by studying the kinematics of K and M giant stars in the
solar neighbourhood\ (Famaey et al. 2004).

The present contribution is organized as follows. 
Sect. 2 briefly describes the data recently published by Famaey 
et al.\ (2004). The salient conclusions of the kinematic analysis of these
data are  presented in Sect.~3, and Sect.~4 confronts the results with
Eggen's hypothesis. We then argue in favour of the alternative concept of
dynamical streams and discuss its consequences for the dynamics of the
Galaxy in Sect.~5. A possible relationship with the frequency of
stars hosting exoplanets in the solar neighbourhood is suggested in
Sect.~6, while we refer to Sect.~7 for the wonderful insight that GAIA
will offer into this subject.

\section{Data}

\begin{table*}[htb]
  \caption{Kinematic parameters of the different base groups identified by Famaey et al.\ (2004): the mean velocities $U_0$, $V_0$ and $W_0$ of each subgroup in the cartesian coordinate frame centered on the sun, as well as the velocity dispersions $\sigma_{U'}$, $\sigma_{V'}$ and $\sigma_W$ along the axes of the velocity ellipsoid, and the vertex deviation $l_v$. The probability that a given star belongs to a given subgroup has been calculated using Bayes formula. The maximum probability then defines the subgroup to which the star is assigned. The number $n$ of stars in each subgroup is given in the last row. For all the details, refer to Famaey et al.\ (2004).}
  \label{}
  \begin{center}
    \leavevmode
      \begin{tabular}{l| l | l | l | l | l | l }
\hline
  & Young group & Hot group & Hyades stream & Sirius stream & Hercules stream & Background \\
\hline
$U_0$ (\kms) & -10.41 $\pm$ 0.94 & -18.50 $\pm$ 2.82  & -30.34
  $\pm$ 1.54 & 6.53 $\pm$ 1.93 & -42.13 $\pm$ 1.95 & -2.78 $\pm$ 1.07 \\
$\sigma_{U'}$ (\kms) & 15.37 $\pm$ 1.09 &  58.02 $\pm$ 1.87 & 11.83 $\pm$
  1.34 & 14.37 $\pm$ 2.05 & 28.35 $\pm$ 1.68 & 33.30 $\pm$ 0.70 \\
$V_0$ (\kms) & -12.37 $\pm$ 0.90 & -53.30 $\pm$ 3.10 & -20.27 $\pm$
  0.62 & 3.96  $\pm$ 0.67 & -51.64  $\pm$ 1.07 & -15.42 $\pm$ 0.82\\
$\sigma_{V'}$ (\kms) & 9.93 $\pm$ 0.75 & 41.36 $\pm$ 1.70 & 5.08 $\pm$
  0.76 & 4.63 $\pm$ 0.71 & 9.31 $\pm$ 1.22 & 17.94 $\pm$ 0.80\\
$W_0$ (\kms) & -7.75 $\pm$ 0.57 & -6.61 $\pm$ 1.82 & -4.82 $\pm$
  0.80 & -5.80 $\pm$ 1.15 & -8.06 $\pm$ 1.30 &  -8.26 $\pm$ 0.38 \\
$\sigma_W$ (\kms) & 6.69 $\pm$ 0.58 & 39.14 $\pm$ 1.69 & 8.75 $\pm$
  0.74 & 9.66 $\pm$ 0.82 & 17.10 $\pm$ 1.63 & 17.65 $\pm$ 0.34 \\
$l_v$ ($^\circ$) & 16.40 $\pm$ 10.26 & 0.16 $\pm$ 4.77 & -8.79 $\pm$
  4.05 & -11.96 $\pm$ 3.53 & -6.53 $\pm$ 2.77 & -2.18 $\pm$ 1.60 \\
$n$ & 413 & 401 & 392 & 268 & 529 & 4027\\

\hline

\end{tabular}

  \end{center}
\end{table*}

Fifteen years of observations with the CORAVEL spectrometer on the Swiss 1-m telescope at the {\it Observatoire de Haute Provence} yielded the radial velocities of Hipparcos stars later than about F5 in the northern hemisphere.    
This unique database, comprising about 20\,000 stars 
(all with $\delta > 0^\circ$) measured with a typical precision of
0.3~\kms, combines a  high precision and the absence of kinematic bias.
The radial velocities of 6030 K and M giants, completing the Hipparcos
parallaxes and the Tycho-2 proper motions, were recently published by
Famaey et al.\ (2004). Six-dimensional phase-space data are thus
available for this sample, as it will be the case with GAIA. The present
contribution thus illustrates the kind of studies that may be repeated 
on a much larger scale with GAIA, thus opening new perspectives (see
Sect.~7). 

\section{Kinematic analysis}

In order to analyze the kinematics of the sample briefly described in Sect. 2, Famaey et al.\ (2004) have assumed that it is a mixture of several base groups having simple gaussian distributions both in velocity space and in luminosity. This is obviously not completely rigorous, but has the advantage of enabling us to use a bayesian method to identify and quantify the different subgroups present in the data and possibly related to extremely complex dynamical phenomema, which cannot be easily parametrized. 

The parameters of the different base groups were derived using the maximum-likelihood method of Luri et al.\ (1996). The best stable solution found was a solution with six base groups. All the kinematic parameters of these groups are listed in Table 1. First, a group of young and luminous stars is concentrated near the origin of velocity space, with a vertex deviation of $16.4^\circ$. A hot group probably composed  of halo and thick disk stars is also present. Three streams were found, one corresponding to the Hercules stream (composed of stars lagging behind the galactic rotation and moving outward), and the other two corresponding to the famous Hyades and Sirius superclusters. Finally, a background group with no vertex deviation was found to represent 60\% of the stars. 

The most surprising property of this background group is that it  is not centered on the value commonly accepted for the antisolar motion:  it is centered instead on $U_0 = -2.78\pm1.07$~\kms.
The young group is however centered on the commonly
accepted value of $U_0 =-10.41\pm0.94$~\kms\ and this difference clearly prevents us from deriving without ambiguity the solar motion.
This discrepancy raises the essential question of how to derive the
solar motion:  does there exist in the disk a subset of stars having no net radial motion? If the smooth background is indeed an axisymmetric background with no net radial motion, we have found a totally new value for the solar motion. Nevertheless, we have no strong argument to assess that this is the case, especially if the  streams have a dynamical origin, as suggested in the next sections. An analysis of the radial velocities of stars very near Sagittarius A$^*$ could well answer the question of the radial motion of the Sun in the Galaxy, but the current uncertainties on those velocities and on the distance to the Galactic centre are still too large to yield reliable results. Perhaps we shall have to wait until the GAIA mission to have this question answered.

Thanks to the maximum likelihood method\ (Luri et al. 1996), Famaey et al.\ (2004) have derived unbiased absolute magnitudes for the stars of the different base groups (taking into account extinction and the parallax uncertainties). Comparing their location  in the Hertzsprung-Russell diagram with isochrones reveals a very wide range of ages for the stars belonging to the superclusters, ranging from several hundreds Myrs to several Gyrs. Moreover, the Hyades supercluster seems to be slightly more metal-rich than the background group of stars.

\section{Failure of Eggen's scenario}

In this section, we confront the classical Eggen's scenario with the results of the kinematic study for the Hyades and Sirius superclusters. 
This scenario states that a large number of stars were formed (almost)
simultaneously in a certain region of the Galaxy and created a
cluster-like structure with a well-defined position and velocity. After
several galactic rotations, the cluster partially evaporated and formed a tube called
supercluster. Stars in the supercluster still share common $V$
velocities when located in the same region of the tube (for example the
solar neighbourhood). Indeed, disk stars (most of which move
on quasi-circular epicyclic orbits) which formed at the same place and
time, and which stayed together in the Galaxy after a few galactic
rotations (since they are all currently
observed in the solar neighbourhood) must necessarily have the same period
of revolution around the Galactic center, and thus the same guiding-center and thus the same velocity $V$. This theory has thus the great advantage of predicting extended horizontal branches
crossing the $UV$-plane similar to those observed in Famaey
 et al.\ (2004). Moreover, in this framework, it is easy to account for
the fact   that the Hyades supercluster seems to be more metal-rich than
the smooth background, by assuming that the giant molecular cloud out of
which it was created was a local (metallicity) anomaly. However, to
explain the wide range of ages observed in a given supercluster
(also reported by Chereul \& Grenon 2001 for the Hyades),
the stars must have formed at different epochs out of one and the same
large molecular cloud, keeping the same kinematics and the same
metallicity for several Gyrs. Chereul et al.\ (1998) suggested that the
supercluster-like velocity structure is just a chance juxtaposition of
several cluster remnants, but this hypothesis requires extraordinarily
long survival times for the oldest clusters (with ages $> 2$ Gyr) in the
supercluster-like structure, implying that the primordial clusters were
extraordinarily massive (several $10^4$~M$_\odot$). Moreover, the metal
content of the different cluster remnants must also be the same by
chance. This hypothesis thus requires too much chance coincidence to be
considered as the valid one in order to interpret the supercluster
phenomenology.

\section{Dynamical streams}

As an alternative, dynamical mechanisms  caused by the disturbing
effect of a non-axisymmetric component of the gravitational
potential are a more satisfactory explanation. The Hercules stream was
recently considered responsible for the bimodal character of the local velocity
distribution\ (Dehnen 2000), which is due to the rotation of the bar if the Sun
is located at the bar's outer Lindblad resonance (OLR). Indeed, stars in the Galaxy
will have their orbits elongated  along or perpendicular to the major
axis of the bar (orbits respectively called the LSR and 
OLR modes in the terminology of Dehnen 2000),  depending upon their position relative to the resonances, and
both types of orbits coexist at the OLR radius. This likely
dynamical origin of the Hercules stream has been the first example of a
non-axisymmetric origin for a stream in velocity space: the other streams
could thus be related to other non-axisymmetric perturbations, such as the spiral arms.

Chakrabarty\ (2004) points out that other streams could appear in the vicinity of the sun when a weak spirality is added to the effect of the bar. In this scenario, while the Hercules stream stars are pushed away from the solar neighbourhood to outer regions of the Galaxy, the Hyades stream would be its equivalent, but for stars coming from inner regions of the Galaxy, which would explain the high metal content of the Hyades stream (because the interstellar medium is more metal-rich in the inner regions of the Milky Way). This study points out that a weak spirality alone is not sufficient to create the observed streams.

Nevertheless, De Simone et al.\ (2004) have shown that the structure of the local distribution function
could well be due to a lumpy potential related to the presence of strong transient spiral waves. Besides those simulations, a recent model of gas flows in the Galaxy \ (Bissantz et al. 2003) indicates that the amplitude of the spiral structure in the mass density is larger by a factor $1.5$ than its amplitude in the near-infrared luminosity density. Interestingly, the corotation radius of the spiral pattern in this model is very close to the Sun's position. Thus, the two major low-velocity streams identified by Famaey et al.\ (2004), Hyades and Sirius, can be interpreted as the inward- and outward-moving streams of 
stars on horseshoe orbits that cross corotation (Sellwood \& Binney 2002). In the past, the importance of stirring by spiral structure has been underestimated because it was thought that spirals heated the disk strongly, and because the amount of heating is observationally constrained. However, Sellwood \& Binney\ (2002)  showed that the dominant effect of spirals is to stirr without heating. In this scenario, the peculiar chemical composition of the Hyades stream (i.e., a metallicity higher than average for field giants) suggests that the group has a common galactocentric origin in the inner Galaxy (where the interstellar medium is more metal-rich than in the solar neighbourhood) and that it was pushed in the solar neighbourhood by the spiral wave, while the Sirius stream would have an origin closer to the Sun and would be the counterpart of the Hyades stream, leaving the global angular momentum distribution unchanged. The clusters of coeval stars with which these streams have historically been identified are thus picked up by the spiral gravitational field along with field stars of all ages and
backgrounds. Indeed, the scale of a cluster is small in comparison to the scale of the perturbation of the potential linked with a spiral wave, and the response of a cluster as a whole to a spiral wave is thus similar to the response of a single star.

If this dynamical scenario is correct, the term {\it dynamical stream} for the clustering in
velocity space seems more appropriate than the term {\it supercluster} since
it is not caused by contemporaneous star formation but rather involves stars
that do not share a common place of birth: stars in the streams just share at
present time a common velocity vector.

This revision of our understanding of stellar motions in the Galaxy has dramatic consequences:
\begin{itemize}
\item The radial displacements have to be taken into account if one wants to describe the past evolution of the Galaxy, and this is very hard to do because the signature of these events vanishes rapidly (the observed peculiar motions are recent -- about 100 Myrs). This warning may apply to most stars in our Galaxy as they may be expected to have been dynamically disturbed by a spiral wave at least once in their lifetime.

\item As pointed out in Sect. 3, we cannot be sure that there is no net radial motion for field stars in the solar neighbourhood, which was the basic hypothesis used so far to derive the solar motion.

\item Models of the chemical evolution of the Milky Way (and disk galaxies in general) will have to be radically revised. The radial mixing due to spiral waves could in fact be the key to understanding the presence of an old metal-rich population in the solar neighbourhood (Haywood 2005).

\item Many exoplanetary systems observed in our neighbourhood could 
have been linked with such radial displacement in the past. It is 
well kown\ (e.g. Santos et al. 2003) that stars with planetary systems
are more metal-rich, on average, than "normal" stars. It is thus likely 
that they formed in dense gas clouds with a high metal content, such
as those located in the more central regions of the Milky Way. The
dynamical streams could thus be the mechanism that brought many of them
closer to the Sun.  This suggestion is discussed in more details in
Sect.~\ref{Sect:exoplanets}.

\item The strength of the spiral needed to create such streams, also probed by observations of the radio-frequency lines of H and CO\ (Bissantz et al. 2003), requires the disk to be massive even near the sun. Moreover, the microlensing results show that essentially all the Galactic mass within 4 kpc from the Galactic centre is in stars\ (Bissantz \& Gehrard 2002). So there is really very little room for dark matter inside the solar radius. This seems in conflict with predictions of standard cold dark matter cosmology, and maybe opens the era of the precedence of modified gravity over exotic matter\ (Famaey \& Binney 2004).
\end{itemize}

Nevertheless, there is another alternative to Eggen's scenario. Indeed, recent investigations\ (Navarro et al. 2004) point out that remnants of mergers could be present in the velocity substructure of the disk: it
seems to be the case of the Arcturus group ($U \simeq 0$~\kms, $V \simeq
-115$~\kms). In this scenario, the streams observed in our kinematic study could thus be the
remnants of merger events between our Galaxy and a satellite galaxy. The
merger would have triggered star formation whereas the oldest giant stars
would be stars accreted from the companion galaxy. However, a recent merger with a
satellite galaxy would moreover induce a perturbation in $W$, which is far from clear in the study of Famaey et al.\ (2004; see also Table~1). If the hierarchical ('bottom-up') cosmological
model is correct, the Milky Way system should have accreted and subsequently tidally destroyed approximately 100 low-mass galaxies in the past 12 Gyr,  which leads to
one merger every 120 Myr, but with a decreasing rate. Thus, the chance that two of them
(leading to the Hyades and Sirius streams) have left such noticeable signatures in the disk near the position of the Sun in the last Gyr is statistically unlikely (although  not impossible). 

\section{Are stars hosting exoplanets more frequently found within dynamical streams?}
\label{Sect:exoplanets}

\begin{figure}[!ht]
  \begin{center}
    \leavevmode
\centerline{\epsfig{file=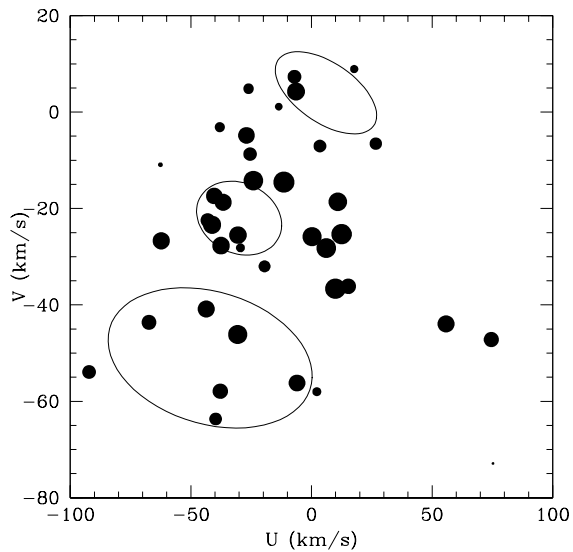, width=0.45\textwidth}}
  \end{center}
  \caption{\label{Fig:exoplanets}
Location  in the ($U,V$) plane of stars hosting exoplanets from the 
CORALIE survey (crosses; Santos et al. 2003). The symbol size is
proportional to [Fe/H].  The Hyades, Sirius and Hercules streams have
been represented by their 1.5$\sigma$ velocity ellipsoids, according to
the parameters listed in Table~1. }
\end{figure}

As suggested in the previous section, it is worth testing whether or not 
stars hosting exoplanets are preferentially found among dynamical streams
like the Hyades one carrying metal-rich stars. To this end,
Fig.~\ref{Fig:exoplanets} compares the location in the ($U,V$) plane of
stars with exoplanets from the CORALIE survey (Santos et al. 2003) with
the location of the streams found by Famaey et al. (2004). Note that the
$U$ and $V$ velocities from Santos et al. (2003) have been  uncorrected
for the solar motion, using the same values as in that study, namely
$U_\odot = 10$~\kms\
and $V_\odot = 6$~\kms. The comparison has been
restricted to the 41 stars with exoplanets from the CORALIE survey to
avoid any kinematical bias.

Clearly, there is a clump of stars with exoplanets falling among the 
Hyades stream. To be more quantitative, the number of exoplanet stars in
the $1.5\sigma$ ellipsoid associated with the Hyades stream has been
compared to the corresponding number for K and M giant stars. A binomial
test on the data  listed in Table 2 yields the following result: given $p
= 0.08,\; q = 0.92$ from the sample of giant stars, the expected number of
exoplanet stars in the Hyades stream would be $ \langle n \rangle =
N_{\rm tot}\; p = 3.18$, with a dispersion $(N p q)^{1/2} = 1.71$  (where
$N_{\rm tot} = 41$ is the total number of exoplanet stars) if they were
extracted from the same parent distribution.  Given that 7 stars are
actually observed (instead of 3.18$\pm1.71$), there is a clear excess (at
a 2.2~$\sigma$ level) of stars hosting exoplanets in the Hyades stream. 

\begin{table}
\caption{
Number of stars within the 1.5$\sigma$ velocity ellipsoid of the Hyades 
stream, for stars hosting exoplanets and for  K and M giants.
}
 \begin{center}
    \leavevmode
\begin{tabular}{llllll}
\hline\\[-5pt]
         &\multicolumn{2}{c}{Exoplanet stars} &&
\multicolumn{2}{c}{K and M giants}\\
$N_{\rm tot}$&\multicolumn{2}{c}{41} &&
\multicolumn{2}{c}{6030}
\\[+5pt]
\cline{2-3}\cline{5-6}\\
 & $N$ & $N/N_{\rm tot}$ &&$N$ & $N/N_{\rm tot}$ \\
\hline\\[-5pt] 
Hyades &  7 & 0.17 && 467 & 0.08 \\
\hline\\
\end{tabular}
\end{center}
\end{table}

\section{What is to be expected from GAIA?}

The kind of study as described in the present paper will largely benefit from GAIA. First,  all the necessary data (including radial velocities and metallicities) will be delivered by GAIA, with no need for tedious ground-based follow-up observations. Second, the large galactic volume probed by GAIA will make it easy to answer several exciting questions raised by our identification of dynamical streams in the solar neighbourhood; namely:
\begin{itemize}
\item What is the spatial extent of dynamical streams?
\item Are similar streams present in other parts of the disk, or is the solar neighbourhood (or rather the solar galactocentric distance) special in that respect?
\end{itemize}

Finally, the most important goal of the GAIA mission for astronomy and physics as a whole is of course that the case for cold dark matter in the disk and the halo will be thoroughly investigated (see e.g.  Binney 2005 for a review).

\section*{Acknowledgments}
We are grateful to F. Pont and Professor J. J. Binney for interesting 
discussions and suggestions. We also thank G. Traversa and B. Pernier for
their important contribution to the observations with the CORAVEL
spectrometer installed on the Swiss telescope at the {\it Observatoire de
Haute Provence}, operated with grants from the {\it Fonds National suisse
de la Recherche Scientifique}. B.F. is post-doctoral fellow from the
{\it Fondation Wiener-Anspach} (Belgium). A.J. is Senior Research
Associate from the {\it Fonds National de la Recherche Scientifique}
(Belgium).

\end{document}